\documentclass[aps,pra,twocolumn,amsmath,amssymb,showpacs,superscriptaddress,tightenlines]{revtex4}
\usepackage{graphicx}% Include figure files
\usepackage{epsfig}
\usepackage{dcolumn}% Align table columns on decimal point
\usepackage{bm}% bold math

\usepackage{amssymb}
\usepackage{bm}
\usepackage{amsmath}
\usepackage[colorlinks,citecolor=blue,linkcolor=blue,hyperindex,CJKbookmarks,dvipdfm]{hyperref}
\begin{document}

\title{Photon emission via vacuum-dressed intermediate states under ultrastrong
coupling}

\author{Jin-Feng Huang}

\address{Department of Physics and Institute of Theoretical Physics, The Chinese
University of Hong Kong, Shatin, Hong Kong Special Administrative
Region, People's Republic of China}

\author{C. K. Law}

\address{Department of Physics and Institute of Theoretical Physics, The Chinese
University of Hong Kong, Shatin, Hong Kong Special Administrative
Region, People's Republic of China}

\date{\today}
\begin{abstract}

We investigate theoretically quantum effects of a cavity-atom
system in which the upper two levels of a cascade-type three-level
atom interact with a cavity field mode in the ultrastrong coupling
regime. By exploiting the virtual photons carried by atom-cavity
dressed states, we indicate how two external driving fields induce
Raman transitions such that a continuous conversion of virtual
photons into real photons can be achieved with a high probability. We present analytical and numerical solutions of the system. In
addition, we show that the converted photons exhibit a bunching
behavior in the steady state. Our scheme demonstrates that
atom-cavity dressed states in the ultrastrong coupling regime can
serve as intermediate states to invoke nonlinear optical
processes.
\end{abstract}
\maketitle

\section{Introduction}

Quantum behavior of a two-level system (atom) interacting with a
single-mode electromagnetic field in the ultrastrong coupling
regime has been a subject of research interest recently. Such a
regime is signaled by the breakdown of rotating wave approximation
(RWA) because the vacuum Rabi frequency is high enough to be an
appreciable fraction of cavity field frequency $\omega_{c}$ or
atomic transition frequency $\omega_{A}$, and this could be
realized by superconducting qubits in microwave
resonators~\cite{Blais2009PRA,Mooij2010PRL,Gross2010NatPhy}, and
by intersubband transitions in semiconductor microcavities
~\cite{Beltram2009PRB,Huber2009Nat,Sirtori2010PRL}. Furthermore,
an ultrastrong coupling Hamiltonian could be simulated by cavity
QED with natural atoms~\cite{Parkins2013PRA} and various circuit
QED~\cite{SolanoPRX2012} settings.

For an atom-cavity system under ultrastrong coupling,
counter-rotating terms in the Hamiltonian can significantly modify
the energy states of the system \cite{Braak}. Quantum transitions
between different energy states can lead to the asymmetry of
vacuum Rabi-splitting~\cite{Cao2011NJP}, superradiance
transition~\cite{Ashhab2013PRA}, and non-classical photon
statistics~\cite{Hartmann2012PRL,Hartmann2013PRL}. In addition,
counter-rotating terms can affect quantum dynamics, such as the
collapse and revival behavior~\cite{Solano2010PRL}, quantum Zeno
effect~\cite{Cao2010PRA,Ai2010PRA}, single photon
scattering~\cite{Wang2012PRA}, and collective spontaneous emission in multi-atom systems~\cite{Scully2009PRL,Rohlsberger2010Sci,Li2013PRA}.

Physically, counter-rotating terms correspond to energy
non-conserving processes involving virtual photons. In order to
study such virtual photons directly, it is interesting to convert
the virtual photons into real photons by suitable designs of the
Hamiltonians. For example, by manipulating the time-dependence of
the atom-field coupling
strength~\cite{Huber2009Nat,Liberato2009PRA,Garziano}, or by
introducing an atomic decaying channel through a $\Xi$-type
three-level atom~\cite{Savasta2013PRL}. In the latter case, Stassi
\textit{et al.} have considered a configuration in which the upper
two levels of a $\Xi$-type atom ultrastrongly couples to a cavity
mode. Once the atom is initially prepared in one of the upper two
levels, the decay to the ground level is accompanied by the
emission of two real photons \cite{Savasta2013PRL}. A similar
$\Xi$-type configuration was proposed by Carusotto \textit{et al.}
who investigated the back-reaction effects of quantum vacuum when
the system is continuously driven by an external field of a single
frequency~\cite{Carusotto2012PRA}.

In this paper we examine the quantum dynamics of the $\Xi$-type
configuration in the presence of two external driving fields of
different frequencies. Different from the models considered in
Refs.~\cite{Savasta2013PRL,Carusotto2012PRA}, the two external
driving fields in our scheme play the role of pump and Stokes
fields such that they can coherently induce Raman transitions with
the vacuum-dressed upper two levels as intermediate states. In
this paper we study how real photons can be generated by
exploiting virtual photons in the intermediate states. As we shall
discuss below, the Raman configuration has a key advantage that a
continuous photon emission can be achieved, and the emission rate
can be substantially enhanced by the Stokes driving field. In
addition, as a result of the Raman configuration, we show that a
dark state, which is composed by the superposition of cavity
vacuum and a cavity photon pair, occurs.

Our paper is organized as follows. In Sec. II, we introduce the
model Hamiltonian. Then in Sec. III, we present analytical and
numerical solutions of non-dissipative quantum dynamics under far
off resonance and resonance conditions. In particular, we indicate
the conditions to reach an almost deterministic conversion of a
virtual photon pair into a real photon pair. In Sec. IV, we
address the effects of cavity field damping and atomic decay by
using the master equation approach. By the solutions of the
density matrix we study the output cavity photon rate and the
second order coherence function. Finally, we draw our conclusions
in Sec. V.

\section{Hamiltonian}

\begin{figure}
\includegraphics[bb=24bp 524bp 540bp 797bp,clip,scale=0.45]{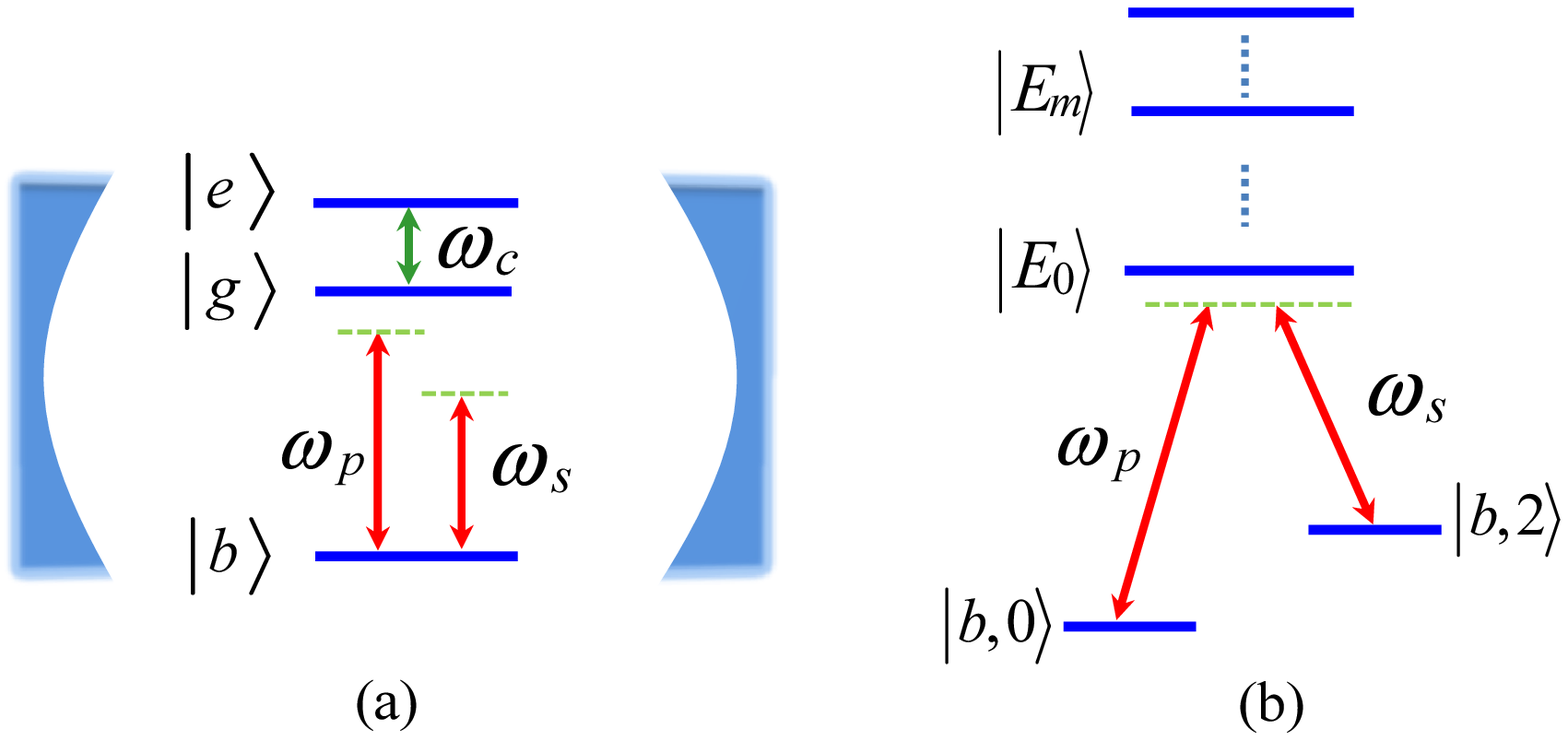}\caption{\label{fig:1}(Color online)
(a) Schematic of the system in which a $\Xi$-type three-level atom is
placed in a cavity.  The atom is driven by two external fields of
frequency $\omega_p$ and $\omega_s$. The upper two levels
$\left|e\right\rangle $ and $\left|g\right\rangle $ of the atom
resonantly couple to a single cavity mode. (b) Energy level
diagram of the system and the Raman type transition. }
\end{figure}

The system under investigation consists of a cascade-type
three-level atom confined in a single-mode cavity with the upper
two levels $|e\rangle$ and $|g\rangle$ resonantly coupled to the
cavity field mode (Fig. 1). The energy difference between the
bottom level $|b\rangle$ and the middle level $|g\rangle$ is
assumed to be sufficiently different from the cavity field
frequency such that the cavity field does not interact the atom in
the state
$|b\rangle$~\cite{Savasta2013PRL,Carusotto2012PRA,switch}. We
remark that in circuit QED, physical properties of three-level
artificial atoms have been discussed theoretically \cite{YouNori},
and demonstrated in experiments \cite{3level_1,3level_2,3level_3}.

In this paper we consider that the system is initially prepared in
the atom state $|b\rangle$ and vacuum cavity field, i.e.,
$|\Psi(0)\rangle=|b,0\rangle$. Then by applying two external
driving fields of frequencies $\omega_{p}$ and $\omega_{s}$,
$\left|b\right\rangle $ and $\left|g\right\rangle $ are coupled.
Our task is to investigate the photon emission process when the
cavity-atom interaction is in the ultrastrong coupling regime.

Specifically, the Hamiltonian of our system is given by
\begin{equation}
H_{S}=H_{0}+H_{D},\label{eq:H}
\end{equation}
where $H_{0}$ is the part in the absence of driving fields, and $H_{D}$
describes the interaction by the two external driving fields ($\hbar=1$):
\begin{eqnarray}
H_{0} & = & \omega_{b}\left|b\right\rangle \left\langle b\right|+\omega_{e}\left|e\right\rangle \left\langle e\right|+\omega_{g}\left|g\right\rangle \left\langle g\right|+\omega_{c}a^{\dagger}a\nonumber \\
 &  & +\lambda\left(a+a^{\dagger}\right)(\left|e\right\rangle \left\langle g\right|+\left|g\right\rangle \left\langle e\right|),\\
H_{D} & = &
(\Omega_{p}\cos\omega_{p}t+\Omega_{s}\cos\omega_{s}t)\left(\left|b\right\rangle
\left\langle g\right|+\left|g\right\rangle \left\langle
b\right|\right).\label{eq:HD}
\end{eqnarray}
Here, $\omega_{i}$ $(i=b,e,g)$ is atomic frequency for level
$|i\rangle$, $\omega_{c}$ is the cavity field frequency, $\lambda$
is the atom-cavity interaction strength, and $\Omega_{p}$ and
$\Omega_{s}$ (the subscripts \textit{p} and \textit{s} label the
pump and Stokes fields) describe the interaction strengths of the
two driving fields. For simplicity, we have assumed that these two
driving fields are in the same phase.

Noting that $H_{R}\equiv H_{0}-\omega_{b}\left|b\right\rangle
\left\langle b\right|$ is the Hamiltonian of the quantum Rabi
model \cite{Rabi}, then by using the eigenvectors $|E_{n}\rangle$
of $H_{R}$, i.e., $H_{R}|E_{n}\rangle=E_{n}|E_{n}\rangle$, $H_{0}$
has the diagonal representation:
\begin{equation}
H_{0} =  \sum_{n=0}^{\infty}E_{n}\left|E_{n}\right\rangle
\left\langle
E_{n}\right|+\sum_{n=0}^{\infty}\left(\omega_{b}+n\omega_{c}\right)\left|b,n\right\rangle
\left\langle b,n\right|.\label{eq:H0}
\end{equation}
Here, $|E_{n}\rangle$ can be expanded as:
$|E_{m}\rangle=\sum_{n=0}^{\infty}c_{mn}|g,n\rangle+\sum_{n=0}^{\infty}d_{mn}|e,n\rangle$
with the real coefficients $c_{mn}\equiv\left\langle
E_{m}|g,n\right\rangle $ and $d_{mn}\equiv\left\langle
E_{m}|e,n\right\rangle $, which can be obtained numerically. In
this way, the driving Hamiltonian (\ref{eq:HD}) becomes:
\begin{eqnarray}
H_{D} & = & \sum_{n,m=0}^{\infty}\Lambda_{mn}(t)\left(\left|b,n\right\rangle \left\langle E_{m}\right|+\left|E_{m}\right\rangle \left\langle b,n\right|\right),\label{eq:HSI}
\end{eqnarray}
where $\Lambda_{mn}(t)\equiv(\Omega_{p}\cos\omega_{p}t+\Omega_{s}\cos\omega_{s}t)c_{mn}$
is defined.

Among the many transitions described in Eq.~(\ref{eq:HSI}), it is
possible to select certain resonant transitions by tuning the
frequencies of the two driving fields. In this paper we shall
focus on the Raman resonance condition
\begin{equation}
\omega_{p}-\omega_{s}=2\omega_{c},\label{eq:twophoRes}
\end{equation}
such that an initial ground state $|b,0\rangle$ can be resonantly
coupled to $|b,2\rangle$ via the intermediate state
$|E_{0}\rangle$ and some higher states $|E_{m}\rangle$ (Fig.~\ref{fig:1}(b)). As long as the driving fields are sufficiently
weak, the system can only access the resonantly coupled states
effectively. As we shall discuss in Sec. III, $H_{D}$ is well
approximated by keeping $n=0$ and $n=2$ terms.

\begin{figure}
\includegraphics[width=2.8 in]{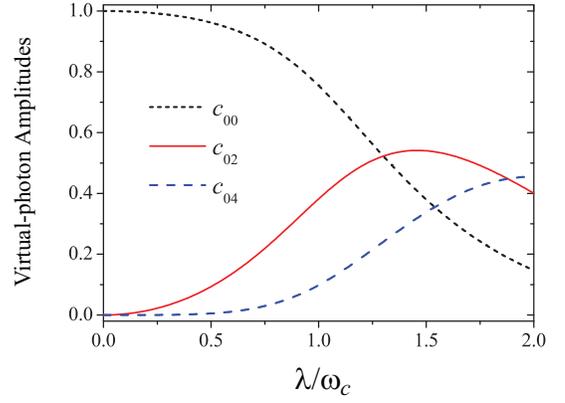}
\caption{\label{fig:2}(Color online) Probability amplitude of
$\left|g,0\right\rangle $, $\left|g,2\right\rangle $ and
$\left|g,4\right\rangle $ in the ground state of $H_{R}$ as a
function of the coupling strength $\lambda$ for the resonant case
$\omega_{0}=\omega_{c}$.}
\end{figure}

It is important to note that if $\lambda$ is not strong enough,
i.e., the system is not in the ultrastrong coupling regime, then
$H_{R}$ is reduced to the Jaynes-Cummings (JC) Hamiltonian by RWA.
In this case $|E_{0}\rangle\approx|g,0\rangle$ is the only
(intermediate) state that couples to $|b,0\rangle$ by the driving
fields, and since the driving fields do not change the photon
number, $|g,0\rangle$ cannot be coupled to $|b,2\rangle$. In other
words, the Raman transition from $|b,0\rangle$ to $|b,2\rangle$ is
impossible in the JC regime. On the other hand, since
$|E_{0}\rangle=
c_{00}|g,0\rangle+c_{02}|g,2\rangle+c_{04}|g,4\rangle+d_{01}|e,1\rangle+... $ in the
ultrastrong coupling regime, with $c_{02}$ being significant, this
makes the Raman transition $|b,0\rangle$ to $|b,2\rangle$
possible. In Fig. \ref{fig:2} we show how $c_{02}$ increases as
$\lambda$ increases. When the coupling is in the deep strong
regime \cite{Solano2010PRL}, $\lambda/\omega_{c} > 1$, $c_{02}$
and $c_{04}$ even become larger than $c_{00}$.

\section{Non-dissipative quantum dynamics}

In order to indicate the essential features of our interaction
scheme, we first discuss the quantum dynamics in the absence of
damping. The damping effects will be addressed in the next
section.

\subsection{Large detuning case}

In the interaction picture with respect to $H_{0}$, the system
Hamiltonian reads:
\begin{eqnarray}
H_{D}\left(t\right) & = & \sum_{n,m=0}^{\infty}\sum_{q=\pm1}\sum_{l=p,s}
\Omega_{l,mn}\left|E_{m}\right\rangle \left\langle b,n\right|e^{i\delta_{mn,ql}t}\nonumber \\
 &  & +{\rm {H.c.},}\label{eq:Vapp}
\end{eqnarray}
where $\Omega_{l,mn}$ and the detuning $\delta_{mn,ql}$ are
defined by,
\begin{eqnarray}
\Omega_{l,mn} & = & \frac{\Omega_{l}}{2}c_{mn}\quad\quad\left(l=p,s\right),\nonumber \\
\delta_{mn,ql} & = & E_{m}-\omega_{b}-n\omega_{c}+q\omega_{l}.
\end{eqnarray}
We focus on the large detuning regime defined by the conditions:
$E_{0}-\omega_{b}-\omega_{p}\gg\Omega_{p}c_{00}/2$, and
$E_{0}-\omega_{b}-2\omega_{c}-\omega_{s}\gg\Omega_{s}c_{02}/2$,
which allow us to adiabatically eliminate the energy levels
$\left|E_{m}\right\rangle $ $(m=0,1,2,..)$. Consequently,
$H_{D}(t)$ under the Raman resonance condition
(\ref{eq:twophoRes}) can be approximated by an effective
Hamiltonian:
\begin{eqnarray}
H_{eff} & \cong & \sum_{n}^{\infty}\Delta_{n}\left|b,n\right\rangle \left\langle b,n\right|+\sum_{n=0}^{\infty}g_{n,n+2}\left|b,n+2\right\rangle \left\langle b,n\right|\nonumber \\
 &  & +{\rm \mbox{H.c.},}\label{eq:Heff-2}
\end{eqnarray}
where the AC Stark shifts $\Delta_{n}$ and effective coupling strength $g_{n,n+2}$ are given by
\begin{eqnarray}
g_{n,n+2} & = & -\sum_{m=0}^{\infty}\Omega_{p,m,n}\Omega_{s,m,n+2}^{*}\left(\frac{1}{\delta_{mn,1s}}+\frac{1}{\delta_{mn,-1p}}\right), \nonumber  \\
&& \\ \Delta_{n} & = &
-\sum_{m=0}^{\infty}\sum_{q=\pm1}\sum_{l=p,s}\frac{\left|\Omega_{l,m,n}\right|^{2}}{\delta_{mn,ql}}.\label{eq:Delta_n}
\end{eqnarray}
In deriving Eq.~(\ref{eq:Heff-2}), we have made use of the time
averaging procedure in Ref.~\cite{Jerke2007CJP}, and we have
discarded some fast oscillating terms, such as
$\left|b,n+2k\right\rangle \left\langle b,n\right|$ ($k\geq2$).
This is justified because the detuning difference in the Raman
process $\left|b,n\right\rangle
\rightarrow\left|E_{m}\right\rangle
\rightarrow\left|b,n+2k\right\rangle $ ($k\geq2$) is much larger
than that in the Raman process $\left|b,n\right\rangle
\rightarrow\left|E_{m}\right\rangle
\rightarrow\left|b,n+2\right\rangle $.

\begin{figure}
\includegraphics[bb=0bp 31bp 424bp 376bp,clip,scale=0.5]{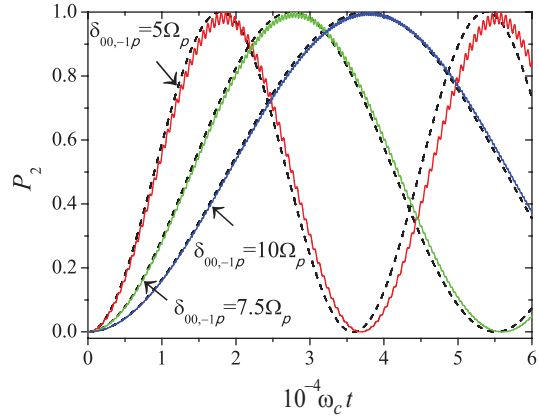}\caption{\label{fig:3}(Color online)
Two-photon probability $P_{2}$ obtained from effective dynamics
Eq.~(\ref{eq:Heff}) (black dashed curve) and exact numerical
calculation (solid curve) for off-resonance cases with detunings:
 $\delta_{00,-1p}=5\Omega_{p}$ (red), $7.5\Omega_{p}$ (green) and $10\Omega_{p}$ (blue).
The ratio of driving strengths used is given by
Eq.~(\ref{eq:ratio1}). Other parameters are
 $\Omega_{p}=2\times10^{-3}\omega_{c}$, $\omega_{s}=\omega_{p}-2\omega_{c}$,
$\omega_{b}=-5\omega_{c}$, $\omega_{e}=\omega_{c},$
$\omega_{g}=0$, $\lambda=0.5\omega_{c}$.}
\end{figure}

A further simplification of Eq.~(\ref{eq:Heff-2}) can be made in
the parameter range $0<\lambda/\omega_{c}<1$, where $c_{04}\ll
c_{02}$ (Fig.~\ref{fig:2}). In such a range of parameter
$g_{2,4}\ll g_{0,2}$ and hence the transition from $|b,2\rangle$
to $|b,4\rangle$ can be neglected as long as the time $t$ we
concern satisfying $g_{2,4}t\ll1$. In this way, we can safely
truncate the effective Hamiltonian~(\ref{eq:Heff-2}) as
\begin{eqnarray}
H_{eff}^{\prime} & = & \Delta_{0}\left|b,0\right\rangle
\left\langle b,0\right|+\Delta_{2}\left|b,2\right\rangle \left\langle b,2\right|
+g_{0,2}\left|b,2\right\rangle \left\langle b,0\right|\nonumber \\
 &  & +{\rm {H.c.}}.\label{eq:Heff}
\end{eqnarray}
Therefore, the system is reduced to a two-level system in which
Rabi oscillations between $|b,0\rangle$ and $|b,2\rangle$ occur.
In particular, the population of $|b,0\rangle$ can be completely
transferred to $|b,2\rangle$ at time $t=\pi/(2g_{0,2})$ when
$\Delta_{0}=\Delta_{2}$. This corresponds to the generation of two
(free) photons, since the atomic state $\left|b\right\rangle $
does not couple to the cavity field mode. Physically, the
two-photon emission is due to the presence of virtual photons in
the intermediate state $\left|E_{0}\right\rangle $ in which the
$c_{02}\left|g,2\right\rangle $ term provides the transition
matrix element. So the photon emission process can be used as a
probe of virtual photons in the vacuum-dressed state of the
atom-cavity system.

We point out that the condition $\Delta_{0}=\Delta_{2}$
corresponds to the balance of AC Stark shifts, and this can be
achieved by choosing a suitable strength ratio of driving fields.
Specifically, from Eq.~(\ref{eq:Delta_n}), $\Delta_{0}=\Delta_{2}$
can be achieved by setting the ratio of driving field couplings at
\begin{equation}
\frac{\Omega_{s}}{\Omega_{p}}=
\sqrt{-\frac{F\left(\omega_{p}\right)}{F\left(\omega_{s}\right)}},\label{eq:ratio1}
\end{equation}
with
\begin{eqnarray}
F\left(x\right) & = & \sum_{m}\left|c_{m2}\right|^{2}\frac{E_{m}-\omega_{b}-2\omega_{c}}{\left(E_{m}-\omega_{b}-2\omega_{c}\right)^{2}-x^{2}}\nonumber \\
 &  & -\sum_{m}\left|c_{m0}\right|^{2}\frac{E_{m}-\omega_{b}}{\left(E_{m}-\omega_{b}\right)^{2}-x^{2}}.
\end{eqnarray}
%\begin{eqnarray}
%F\left(x\right) & = & \sum_{m}\left|c_{m0}\right|^{2}\frac{E_{m}-\omega_{b}}{\left(E_{m}-\omega_{b}\right)^{2}-x^{2}} \nonumber \\
% &  & -\sum_{m}\left|c_{m2}\right|^{2}\frac{E_{m}-\omega_{b}-2\omega_{c}}{\left(E_{m}-\omega_{b}-2\omega_{c}\right)^{2}-x^{2}}.
%\end{eqnarray}

To test the validity of the effective dynamics, we solve
numerically the system evolution based on the original
Hamiltonian~(\ref{eq:H}) and compare it with the prediction by the
effective Hamiltonian~(\ref{eq:Heff}). The results are shown in
Fig.~\ref{fig:3} at the driving field ratio given
by~(\ref{eq:ratio1}). We see that the complete Rabi oscillations
predicted by the effective Hamiltonian match well with numerical
calculations, and the agreement gets better at larger detunings.

It should be noted that the photon emission process here is
different from that in Ref.~\cite{Savasta2013PRL} in which photons
are emitted through the spontaneous decay of $|E_{0}\rangle$,
i.e., $|E_{0}\rangle\rightarrow|b,2\rangle$, assuming the system
is initially prepared at $|E_{0}\rangle$. Since $c_{00}> c_{02}$
in the regime $0<\lambda/\omega_{c}<1$, $|E_{0}\rangle$
predominately decays to $|b,0\rangle$ without emitting photons and
so this would reduce the overall two-photon emission probability
in Ref.~\cite{Savasta2013PRL}. In our scheme, however, a photon
pair can be generated almost deterministically by using the Raman
transition, and $|E_{0}\rangle$ plays only the role of
intermediate state which is rarely populated in the large detuning
regime.

\subsection{Resonance case}

Next we consider the case in which the pump and the Stokes driving
fields are on resonance with the transitions $|b,0\rangle
\leftrightarrow\ensuremath{|E_{0}\rangle }$ and
$\left|E_{0}\right\rangle
\leftrightarrow\ensuremath{\left|b,2\right\rangle }$,
respectively, i.e., $E_{0}-\omega_{b}-\omega_{p}=0$ and
$E_{0}-\omega_{b}-2\omega_{c}-\omega_{s}=0$. In this case, excited
states $|E_{m}\rangle$ ($m>0$) are far off resonance, and hence
they are not effectively involved in the dynamics. Therefore,
(\ref{eq:Vapp}) can be approximated by a $\Lambda$-type
three-level Hamiltonian :
\begin{equation}
H_{eff}^{\prime\prime}=\Omega_{p}^{\prime}\left|E_{0}\right\rangle \left\langle b,0\right|+\Omega_{s}^{\prime}\left|E_{0}\right\rangle \left\langle b,2\right|+{\rm {H.c.},}\label{eq:Vapp1}
\end{equation}
where
\begin{eqnarray}
\Omega_{p}^{\prime} & = & \frac{\Omega_{p}}{2}c_{00},\\
\Omega_{s}^{\prime} & = & \frac{\Omega_{s}}{2}c_{02},\label{eq:Omega_sp}
\end{eqnarray}
are defined. By solving the Schr\"{o}dinger equation governed by
the effective Hamiltonian~(\ref{eq:Vapp1}), it can be shown that
an initial state $\left|b,0\right\rangle $ can coherently evolve
to $\left|b,2\right\rangle $ with the probability
\begin{eqnarray}
P_{2} & = &
\frac{4\eta^{2}\left|c_{00}\right|^{2}\left|c_{02}\right|^{2}}
{\left(\left|c_{00}\right|^{2}+\eta^{2}\left|c_{02}\right|^{2}\right)^{2}}\sin^{4}\frac{\Omega
t}{2},\label{eq:P2_RWA}
\end{eqnarray}
and
$\Omega=\Omega_{p}\sqrt{\left|c_{00}\right|^{2}+\eta^{2}\left|c_{02}\right|^{2}}/2$.
When the strength ratio $\eta = \Omega_s/ \Omega_p$ of the two
driving fields reaches an optimal value $\eta_c$, i.e.,
\begin{equation}
\eta= \eta_c \equiv
\frac{\left|c_{00}\right|}{\left|c_{02}\right|},
\end{equation}
the probability of generating a real photon pair is one at $t=\pi/
\Omega$.

\begin{figure}
\includegraphics[bb=0bp 36bp 460bp 424bp, clip,scale=0.5]{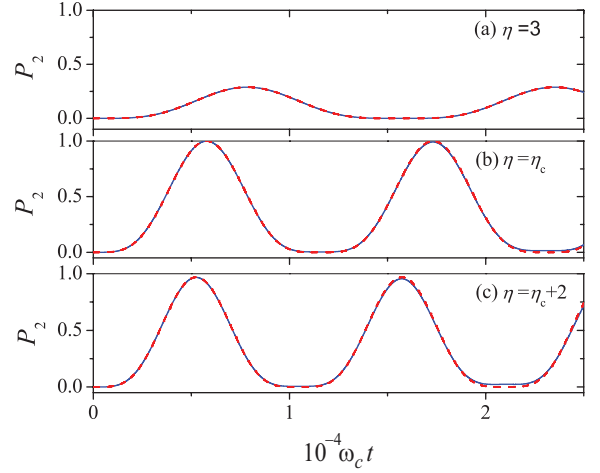}\caption{\label{fig:5}(Color
online) Evolution of two-photon probability $P_2$ for resonance
cases $\omega_{p}=E_{0}-\omega_{b}$ with (a) $\eta=3$, (b) $\eta= \eta_{c}$, and (c) $\eta =
\eta_{c}+2$, where $\eta_{c}=10.2909$,
$\Omega_{p}=0.8\times10^{-3}\omega_{c}$ and
$\lambda=0.5\omega_{c}$. The analytical curve  (red dashed) in Eq.
(\ref{eq:P2_RWA}) and the numerically exact curve (blue solid) are
almost indistinguishable. Other common parameters are
$\omega_{s}=\omega_{p}-2\omega_{c}$, $\omega_{b}=-5\omega_{c}$, $\omega_{e}=\omega_{c},$
$\omega_{g}=0$. }
\end{figure}

To illustrate the validity of the effective three-level dynamics,
we solve numerically the state evolution based on the original
Hamiltonian~(\ref{eq:H}) and compare it with the analytic solution
obtained by the effective Hamiltonian~(\ref{eq:Vapp1}). We show
the probability $P_{2}$ in Fig.~\ref{fig:5} at different driving
ratios $\eta=3,$ $\eta_{c},$ and $\eta_{c}+2$. We see that the
Rabi oscillations predicted by the effective three-level
model~(\ref{eq:Vapp1}) matches well with the exact numerical
solution. In particular $P_2 = 1$ can be achieved at $\eta =
\eta_c$.

\begin{figure}
\includegraphics[bb=0bp 0bp 432bp 347bp,clip,scale=0.5]{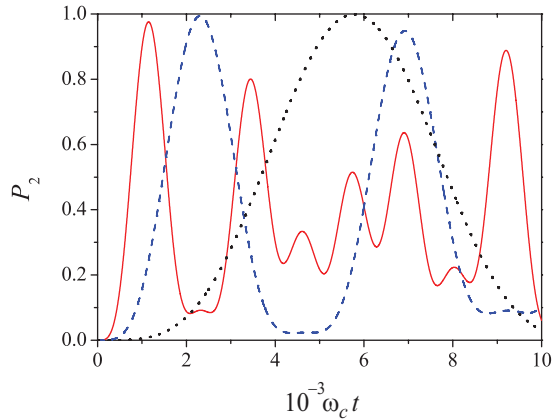}\caption{\label{fig:6}(Color online)
An illustration of the time-dependence of $P_2$ at stronger pump
fields beyond the three-level approximation in Eq.~(\ref{eq:Vapp1}) on resonance $\omega_{p}=E_{0}-\omega_{b}$. The
parameters are: $\Omega_{p}=4\times10^{-3}\omega_{c}$ (red solid
curve), $2\times10^{-3}\omega_{c}$ (blue dashed curve),
$0.8\times10^{-3}\omega_{c}$ (black dotted curve). Other common
parameters are:  $\eta=\eta_{c}$, $\omega_{s}=\omega_{p}-2\omega_{c}$, $\omega_{b}=-5\omega_{c}$,
$\omega_{e}=\omega_{c},$ $\omega_{g}=0$, $\lambda=0.5\omega_{c}$.
}
\end{figure}

We emphasize that the three-level effective
Hamiltonian~(\ref{eq:Vapp1}) is applicable to situations where the
two driving fields are weak, namely
$|\Omega_{l,mn}/(E_{m}-\omega_{b}-n\omega_{c}-\omega_{l})|\ll1$
($m=1,\;2,\ldots$, $l=p,s$). However, for practical purposes, it
is often desirable to speed up the quantum evolution by using
stronger driving fields. To this end, we study the evolution of
$P_{2}$ numerically by the original Hamiltonian~(\ref{eq:H}) at
stronger fields. The results are illustrated in Fig.~\ref{fig:6},
where the driving field strength ratio is kept at $\eta_{c}$. We
see that at stronger driving fields (blue dashed and red solid
curves), $P_{2}$ is no longer described by
Eq.~(\ref{eq:P2_RWA}). This indicates that excitations beyond the
three levels are involved. For the parameters used in
Fig.~\ref{fig:6}, we find that the first maximum of $P_{2}$, which
appears at shorter times for stronger driving fields, is still
close to one even though the three-level approximation breaks
down.

It is interesting to remark that the effective
Hamiltonian~(\ref{eq:Vapp1}) supports a dark state defined by
\begin{equation}
\left|D\right\rangle =\frac{\Omega_{s}^{\prime}}{\sqrt{\Omega_{p}^{\prime2}+
\Omega_{s}^{\prime2}}}\left|b,0\right\rangle -\frac{\Omega_{p}^{\prime}}
{\sqrt{\Omega_{p}^{\prime2}+\Omega_{s}^{\prime2}}}\left|b,2\right\rangle ,
\end{equation}
which satisfies $H_{eff}^{\prime\prime}\left|D\right\rangle =0$.
Such a dark state is a quantum interference effect by which the
atom-cavity system can be decoupled from two external driving
fields. It is a specific coherent superposition of a vacuum state
and a two-photon state while the atom is in the state $|b\rangle$.
Since $\Omega_{s}^{\prime}$ [Eq.~(\ref{eq:Omega_sp})] is
proportional to $c_{02}$, the dark state may be considered as a
signature of the virtual photon pair in $|E_{0}\rangle$.

\section{Effects of damping}

In this section, we address the influence of the cavity field
damping and atomic decay by using the master equation approach.
The damping processes are modelled by system-bath couplings so
that the full Hamiltonian is given by:
\begin{equation}
H=H_{S}+H_{B}+H_{SB},
\end{equation}
where $H_{B}$ and $H_{SB}$ correspond to the bath Hamiltonian and
system-bath interactions respectively,
%where the bath free Hamiltonian reads
\begin{eqnarray}
H_{B}&=&\sum_{\nu=1}^{3}\sum_{\ell}\omega_{\ell}^{\left(\nu\right)}
B_{\ell}^{\left(\nu\right)\dagger}B_{\ell}^{\left(\nu\right)},\nonumber \\
H_{SB} & = &
\sum_{\ell}\alpha_{\ell}^{\left(1\right)}\left(\left|e\right\rangle
\left\langle g\right|+\left|g\right\rangle \left\langle e\right|\right)\left(B_{\ell}^{\left(1\right)\dagger}+B_{\ell}^{\left(1\right)}\right)\nonumber \\
 &  & +\sum_{\ell}\alpha_{\ell}^{\left(2\right)}\left(\left|g\right\rangle \left\langle b\right|+\left|b\right\rangle \left\langle g\right|\right)\left(B_{\ell}^{\left(2\right)\dagger}+B_{\ell}^{\left(2\right)}\right)\nonumber \\
 &  & +\sum_{\ell}\alpha_{\ell}^{\left(3\right)}\left(a+a^{\dagger}\right)\left(B_{\ell}^{\left(3\right)\dagger}+B_{\ell}^{\left(3\right)}\right).
\end{eqnarray}
Here the bath index $\nu$ marks different baths responsible for
different decay channels. $B_{\ell}^{\left(\nu\right)}$
($B_{\ell}^{\left(\nu\right)\dagger}$) is the annihilation
(creation) operator for the $\ell$th mode (with the frequency
$\omega_{\ell}^{(\nu)}$) in the bath $\nu$. Specifically, the
$\nu=1$ and $\nu=2$ baths are respectively for the atomic decay
from $|e\rangle$ to $|g\rangle$ and from $|g\rangle$ to
$|b\rangle$, while the $\nu=3$ bath is for cavity field damping.
These three baths are assumed independent, and
$\alpha_{\ell}^{(\nu)}$ describes the coupling strengths. For weak
system-bath couplings and sufficiently short correlation times of
the baths, we employ the standard Born-Markov
approximation~\cite{Breuer}. In the continuous limit of the baths,
namely making the replacements
$\alpha_{\ell}^{(\nu)}\rightarrow\alpha^{(\nu)}(\omega)$,
$B_{\ell}^{(\nu)}\rightarrow B^{(\nu)}(\omega)$, we obtain the
following master equation at zero temperature~\cite{Blais2011PRA}
\begin{eqnarray}
\frac{d\rho\left(t\right)}{dt} & = & i\left[\rho\left(t\right),H_{S}\right]\nonumber \\
 &  & +\sum_{\nu=1}^{3}\sum_{j,k>j}\Gamma_{kj}^{(\nu)}\left\{ D\left[\left|\varepsilon_{j}\right\rangle \left\langle \varepsilon_{k}\right|\right]\rho\left(t\right)\right\} , \label{eq:Mas_eq}
\end{eqnarray}
with the superoperator $D$ defined as
\begin{equation}
D\left[O\right]\rho\equiv O\rho
O^{\dagger}-\frac{1}{2}O^{\dagger}O\rho-\frac{1}{2}\rho
O^{\dagger}O.
\end{equation}
Here, $|\varepsilon_{j}\rangle$ ($j=1,\;2,\;3,\ldots$) are energy
eigenstates of the system in the absence of driving fields and
damping, i.e.,
$H_{0}|\varepsilon_{j}\rangle=\varepsilon_{j}|\varepsilon_{j}\rangle$.
Therefore $\{|\varepsilon_{j}\}$ is simply
$\{|b,0\rangle,\;|b,1\rangle,\ldots,|E_{0}\rangle,\;|E_{1}\rangle,\ldots\}$
according to Eq.~(\ref{eq:H0}). The relaxation coefficients are
given by:
\begin{equation}
\Gamma_{kj}^{(\nu)}=2\pi|\alpha^{(\nu)}\left(\omega_{kj}\right)|^{2}d^{(\nu)}
\left(\omega_{kj}\right)|C_{jk}^{(\nu)}|^{2}\;(\nu=1,\;2,\;3),
\end{equation}
which depend on the system-bath coupling strength
$\alpha^{(\nu)}(\omega_{kj})$, the spectral density
$d^{(\nu)}(\omega_{kj})$ of bath $\nu$ at the transition frequency
$\omega_{jk}=\varepsilon_{j}-\varepsilon_{k}$, and on the
transition matrix element $C_{jk}^{(\nu)}$:
\begin{eqnarray}
C_{jk}^{(1)} & = & \left\langle \varepsilon_{j}\right|\left(\left|e\right\rangle \left\langle g\right|+\left|g\right\rangle \left\langle e\right|\right)\left|\varepsilon_{k}\right\rangle ,\nonumber \\
C_{jk}^{(2)} & = & \left\langle \varepsilon_{j}\right|\left(\left|g\right\rangle \left\langle b\right|+\left|b\right\rangle \left\langle g\right|\right)\left|\varepsilon_{k}\right\rangle ,\nonumber \\
C_{jk}^{(3)} & = & \left\langle
\varepsilon_{j}\right|\left(a+a^{\dagger}\right)\left|\varepsilon_{k}\right\rangle.
\end{eqnarray}
Note that in writing Eq. (23), we have neglected the (Lamb)
frequency shift terms. For simplicity, we assume
$\alpha^{(\nu)}\left(\omega_{kj}\right)$ and
$d^{(\nu)}\left(\omega_{kj}\right)$ to be
constant~\cite{Savasta2013PRL}. Then the relaxation coefficients
can be written in a compact form
$\Gamma_{kj}^{(\nu)}=\gamma_{\nu}|C_{jk}^{(\nu)}|^{2}=\Gamma_{jk}^{(\nu)}.$
We note that, $C_{jk}^{(1)}$ is non-vanishing only for the
transitions between Rabi levels $\left|E_{m}\right\rangle $.

\subsection{Output photon rate}
\begin{figure}
\includegraphics[width=2.8 in]{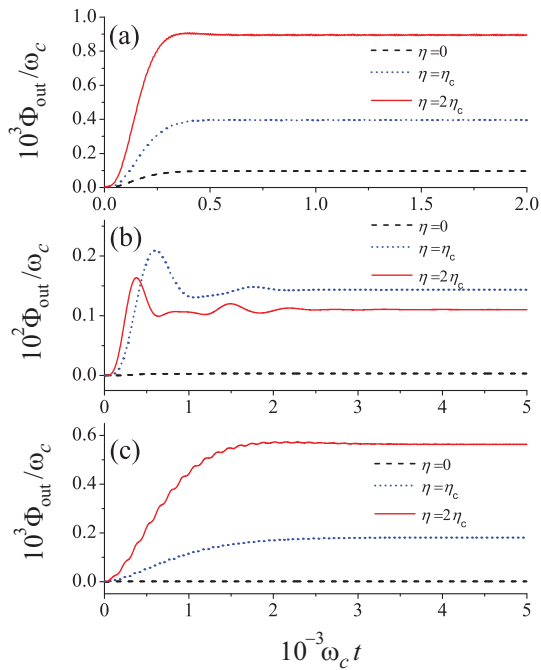}
\caption{\label{fig:7}(Color online) Time dependence of the output
photon rate for $\eta=0$, $\eta_{c}$ and $2\eta_{c}$ with
$\eta_{c}=6.8538$. (a) and (b) are for the resonance
$\omega_{p}=E_{0}-\omega_{b}$ case, and (c) is for an off
resonance  case with $\omega_{p}=4.85\omega_{c}$. The damping
rates are: (a)
$\gamma_{1}=\gamma_{2}=\gamma_{3}=2\times10^{-2}\omega_{c}$, (b)
$\gamma_{1}=\gamma_{2}=\gamma_{3}=2\times10^{-3}\omega_{c}$, (c)
$\gamma_{1}=\gamma_{2}=\gamma_{3}=2\times10^{-3}\omega_{c}$. Other
common parameters used are $\Omega_{p}=8\times10^{-3}\omega_{c}$,
$\lambda=0.6\omega_{c}$, $\omega_{s}=\omega_{p}-2\omega_{c}$,
$\omega_{b}=-5\omega_{c}$, $\omega_{e}=\omega_{c},$
$\omega_{g}=0$. }
\end{figure}

Now we study the features of the output photons released from the
cavity. To this end, we need to adopt a proper generalized
input-output relation in the ultrastrong coupling
regime~\cite{Savasta2013PRL},
\begin{equation}
B_{{\rm {out}}}\left(t\right)=B_{{\rm
{in}}}\left(t\right)-i\sqrt{\gamma_{3}}X^{-}\left(t\right),\label{eq:in_out}
\end{equation}
where the output field operator $B_{{\rm {out}}}(t)$, input field
operator $B_{{\rm {in}}}(t)$ and $X^{-}(t)$ are given by
\begin{eqnarray}
B_{\rm {in}}\left(t\right) & = & \frac{1}{\sqrt{2\pi}}\int d\omega e^{-i\omega\left(t-t_{0}\right)}\left.B^{\left(3\right)}\left(\omega\right)\right|_{t=t_{0}}, \\
B_{\rm{out}}\left(t\right) & = & \frac{1}{\sqrt{2\pi}}\int d\omega
e^{-i\omega\left(t-t_{1}\right)}\left.B^{(3)}(\omega)\right|_{t=t_{1}},
\end{eqnarray}
with $t_{0}\rightarrow-\infty$ and $t_{1}\rightarrow\infty$, and
\begin{eqnarray}
X^{-} & = & \sum_{k,j<k}C_{jk}^{(3)}\left|\varepsilon_{j}\right\rangle \left\langle \varepsilon_{k}\right|,\\
X^{+} & = & \left(X^{-}\right)^{\dagger}.
\end{eqnarray}
Then by the relation~(\ref{eq:in_out}), the output cavity photon
rate is given by
\begin{eqnarray}
\Phi_{{\rm {out}}}\left(t\right) & = & \left\langle B_{{\rm {out}}}^{\dagger}\left(t\right)B_{{\rm {out}}}\left(t\right)\right\rangle \\
 & = & \gamma_{3}\left\langle X^{+}\left(t\right)X^{-}\left(t\right)\right\rangle ,
\end{eqnarray}
assuming the initial field is a vacuum. It is important to note
that, in Eq.~(\ref{eq:in_out}), the operator $X^{-}(t)$ has
replaced $a(t)$ in the standard input-output relations. This is
understood because a photon emission from the cavity is associated
with a transition from a high energy state to a low energy state
defined by $H_S$, and it is described by $X^{-}(t)$. Equivalently,
$X^{-}(t)$ comes from the RWA (i.e., keeping energy conserving
terms) of the system-bath interaction \cite{Hartmann2012PRL}.

In Fig.~\ref{fig:7} we display the time evolution of $\Phi_{{\rm
{out}}}(t)$ for various parameters. For the resonance case, when
the decay rates are much larger than the driving-field strength
$\Omega_{p}$ [Fig.~\ref{fig:7}(a)], namely
$\gamma_{\nu}>\Omega_{p}$ ($\nu=1,\;2,\;3$), the output cavity
photon rate increases from $0$ to a steady value in the long time
limit. This indicates a steady and continuous emission of real
photons. When the decay rates $\gamma_{\nu}$ are in the same order
but smaller than the driving strength $\Omega_{p}$
[Fig.~\ref{fig:7}(b)], namely $\gamma_{\nu}<\Omega_{p}$, the
output cavity photon rate exhibits oscillatory patterns before
settling into steady states. Comparing with Fig.~\ref{fig:7}(a),
we see that the weaker damping rates can increase the steady state
output significantly.  We also plot the output cavity photon rate
for the off resonance case in Fig.~\ref{fig:7}(c). We see how
$\Phi_{{\rm {out}}}\left(t\right)$ reaches a steady value and it
is smaller than that in resonance cases. This is understood
because the large detuning condition (Sec. IIIA) somehow decreases
the effective coupling.
\begin{figure}
\includegraphics[bb=0bp 0bp 346bp 241bp,clip,scale=0.6]{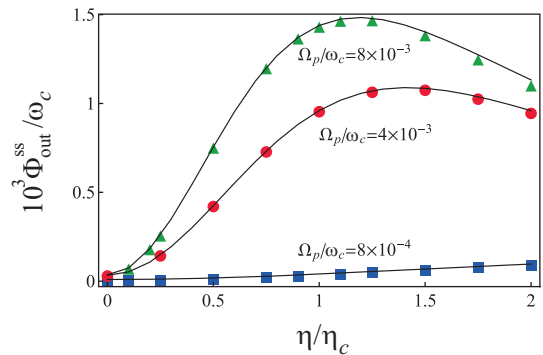}\caption{(Color online) Numerical (discrete points)
and analytical (solid curves) steady output cavity photon rate
$\Phi_{{\rm {out}}}^{ss}$ as a function of the driving strength
ratio $\eta$ under the resonance condition
$\omega_{p}=E_{0}-\omega_{b}$ for various values of $\Omega_{p}$.
Other parameters are the same as in Fig.
\ref{fig:7}(b).}\label{fig:8}
\end{figure}

The effect of Stokes driving field is also illustrated in Fig. 6.
By comparing with the case of $\eta=0$,  we see that the output
photon rate can be enhanced appreciably by the presence of the
Stokes driving field. For example in Fig.~\ref{fig:7}(b), the
magnitude of the enhancement of the output cavity photon rate can
be two orders as a result of the Stokes driving field. In Fig. 7
we plot the dependence of steady state value of $\Phi_{{\rm
{out}}}$ on $\eta$, which indicates that there exits an optimal
value of $\eta$, which is roughly around $\eta_c$ for not too
small $\Omega_p$ (depending on other system parameters), to obtain
a maximum output photon rate in the steady state.

To learn more about the steady state behavior of the system, we
have solved analytically the steady state density matrix
$\rho^{ss}$ based on the effective Hamiltonians~(\ref{eq:Heff})
and~(\ref{eq:Vapp1}). The explicit expressions of $\rho^{ss}$ are
presented in Appendix, and from which the steady state value of
$\Phi_{{\rm {out}}}$ can be obtained as
\begin{eqnarray}
\Phi_{\rm{out}}^{ss} & = &
\gamma_{3}(\rho_{22}^{ss}+2\rho_{33}^{ss}).
\end{eqnarray}
Here the diagonal matrix elements $\rho_{22}^{ss}$ and
$\rho_{33}^{ss}$ [given in (\ref{p1RES}), (\ref{p2RES}), (\ref{P1offres}) and (\ref{P2offres})]
correspond to the probability of $|b,1 \rangle$ and $|b,2 \rangle$
in the steady state. As a check, the analytical
$\Phi_{\rm{out}}^{ss}$ (solid curves in Fig.~\ref{fig:8}) matches
very well with the numerically exact results (discrete points in
Fig.~\ref{fig:8}) for not too strong driving strengths. We remark
that according to the $\rho^{ss}$ solution in Appendix, both
$\rho_{22}^{ss}$ and $\rho_{33}^{ss}$ are proportional to
$\Omega_{l}^{2}$ when damping rates $\gamma_{\nu}$ $(\nu=1,2,3)$
are large compared with driving field strengths $\Omega_{l}$
$(l=p,s)$, i.e., $\gamma_{\nu}\gg\Omega_l$, and so
$\Phi_{\rm{out}}^{ss}$ increases with $\Omega_{l}^{2}$.

\subsection{Photon bunching correlation}

\begin{figure}
\includegraphics[scale=0.425]{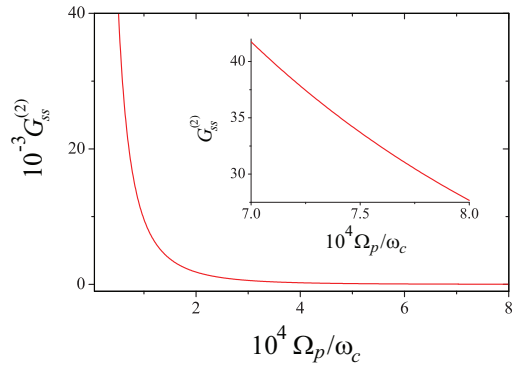}\caption{\label{fig:9}(Color online) Second order
coherence function $G^{(2)}_{ss}$  in the steady state vs the pump
driving strength $\Omega_{p}$ on exact resonance
$\omega_{p}=E_{0}-\omega_{b}$. The parameters used are
$\gamma_{1}=\gamma_{2}=\gamma_{3}=2\times10^{-3}\omega_{c}$,
$\lambda=0.5\omega_{c}$, $\eta=\eta_{c}$,
$\omega_{s}=\omega_{p}-2\omega_{c}$, $\omega_{b}=-5\omega_{c}$,
$\omega_{e}=\omega_{c},$ $\omega_{g}=0$.}
\end{figure}

Finally we examine the correlation between emitted photons in the
steady state. This characterized by the equal-time second order
coherence function $G^{(2)}_{ss}$ defined
by~\cite{Hartmann2012PRL,Savasta2013PRL}
\begin{eqnarray}
G^{(2)}_{ss} & \equiv & \left.\frac{\left\langle
X^{\dagger}\left(t\right)X^{\dagger}\left(t\right)X^{-}\left(t\right)X^{-}\left(t\right)\right\rangle
}{\left\langle
X^{\dagger}\left(t\right)X^{-}\left(t\right)\right\rangle
^{2}}\right|_{t\rightarrow\infty}.
\end{eqnarray}
Again, the cavity mode operator has been replaced by operator
$X^-$. In terms of the approximate steady state density matrix
elements (Appendix), $G^{(2)}_{ss}$ can be expressed as,
\begin{eqnarray}
G^{(2)}_{ss}&  = & \frac{2\rho_{33}^{ss}}{(\rho_{22}^{ss}+2
\rho_{33}^{ss})^{2}}.
\end{eqnarray}
In Fig.~\ref{fig:9} we show the behavior of $G^{(2)}_{ss}$ as a
function of $\Omega_{p}$ at $\eta= \eta_c$. The photon bunching
effect is obviously seen by $G^{(2)}_{ss}>1$ in the figure. In
particular, strong photon bunching $G^{(2)}_{ss} \gg 1$ occurs
when $\Omega_p$ is small compared with the damping rates. This
feature can be described by the approximate density matrix
solution in Appendix, where we find that $G^{(2)}_{ss}\propto
\gamma_{3}^{2}/\Omega_{l}^{2}$ when $\Omega_{l}\ll \gamma_{3}$ for
resonance case, assuming $\gamma$'s are equal.

\section{Conclusion}
To conclude, we have investigated a type of Raman interaction in a
cavity-atom system in which the intermediate states are dressed by
the vacuum field in the ultrastrong coupling regime. By applying a
pump field and a Stokes field, we show how virtual photons in the
intermediate states can be turned into real photons efficiently.
This is shown by the effective Hamiltonian approach for resonance
and far-off resonance cases. Furthermore, the effects of damping
are studied by the master equation, and approximate steady state
solutions are obtained analytically. With the Raman coupling
scheme, photons can be emitted out of the cavity continuously in
the steady state.

Finally, it is worth noting that the photon emission process
described in this paper can be interpreted as a three-photon down
conversion process, since by energy conservation,
$\omega_{p}=\omega_{s}+2\omega_{c}$ in Eq. (6) means a pump photon
of frequency $\omega_p$ is converted into a Stokes photon of
frequency $\omega_s$ and two cavity photons, while the atomic
state remain unchanged. In other words the atom-cavity system in
the ultrastrong coupling regime somehow plays the role of a
nonlinear optical medium, and the interesting source of
nonlinearity comes from the virtual photons in the intermediate
states.

\begin{acknowledgments}
J. F. Huang would like to thank K. M. C. Lee, J. Q. Liao, and L. Jin for their
technical support and T. Y. Li for useful discussions. This work is partially supported by a grant
from the Research Grants Council of Hong Kong, Special
Administrative Region of China (Project No. CUHK401812).
\end{acknowledgments}

\begin{appendix}

\section{Steady state solutions of the master equation}

\subsection{Resonance case}

Under the resonance condition $\omega_{p}=E_{0}-\omega_{b}$ and
$\omega_{s}=E_{0}-\omega_{b}-2\omega_{c}$, and weak driving
fields,
 i.e., $|\Omega_{l,mn}/(E_{m}-\omega_{b}-n\omega_{c}-\omega_{l})|\ll1$
($m=1,\;2,\ldots$, $l=p,s$), the excited states $|E_{m}\rangle$
($m>0$) are far off resonance, and hence they are not effectively
involved in the dynamics. Combined with conditions
$|g_{2,4}|^{2}/\gamma_{3}^{2}\ll1$ and
$|g_{1,3}|^{2}/\gamma_{3}^{2}\ll1$, the system can be
characterized by a three-level $\Lambda$ system~(\ref{eq:Vapp1}).
We define $\rho^{ss}_{n+1,m+1}\equiv\langle
b,n|\rho^{ss}|b,m\rangle$ ($n,m=0,1,2$), and
$\rho^{ss}_{4,4}\equiv \langle E_{0}|\rho^{ss}|E_{0}\rangle$
(superscript ss labels the steady state solution). According to
the effective Hamiltonian~(\ref{eq:Vapp1}), we obtain the steady
state solutions
\begin{eqnarray}
\rho^{ss}_{11} & = & \frac{\Gamma_{21}}{A_{1}}4\left|\Omega_{p}^{\prime}\right|^{2}\left[4\left|\Omega_{s}^{\prime}\right|^{2}\left(\Gamma_{43}-x\right)+x\left(x^{2}+xy+y^{2}\right)\right]\nonumber \\
 &  & +\frac{4\Gamma_{21}}{A_{1}}xy\left|\Omega_{s}^{\prime}\right|^{2}\left[2\left(\Gamma_{41}+\Gamma_{42}+x\right)+\Gamma_{43}\right]\nonumber \\
 &  & +\frac{\Gamma_{21}}{A_{1}}x^{2}y^{2}\left(x+y\right)\nonumber\\
 &  & +\frac{16\Gamma_{21}}{A_{1}}\left|\Omega_{s}^{\prime}\right|^{4}\left(\Gamma_{41}+\Gamma_{42}+x\right)+\frac{16\Gamma_{21}}{A_{1}}x\left|\Omega_{p}^{\prime}\right|^{4}, \nonumber\\
\\
\rho^{ss}_{12} & = & 0,\\
%\end{eqnarray}
%\begin{eqnarray}
 \rho_{13}^{ss} & = & -\frac{4\Omega_{s}^{\prime}\Gamma_{21}\Omega_{p}^{\prime*}}{A_{1}}\left(\Gamma_{41}+\Gamma_{42}+x\right)\left(4\left|\Omega_{s}^{\prime}\right|^{2}+xy\right)\nonumber \\
 &  & -\frac{4\Omega_{s}^{\prime}\Gamma_{21}\Omega_{p}^{\prime*}}{A_{1}}\left[4\left|\Omega_{p}^{\prime}\right|^{2}\left(\Gamma_{43}-x\right)+xy\Gamma_{43}\right], \\
 %\end{eqnarray}
%\begin{eqnarray}
\rho_{14}^{ss} & = & \frac{2ix\Gamma_{21}\Omega_{p}^{\prime*}}{A_{1}}\left[\left(\Gamma_{41}+\Gamma_{42}+x\right)\left(4\left|\Omega_{s}^{\prime}\right|^{2}+xy\right)\right]\nonumber \\
 &  & +\frac{2ix\Gamma_{21}\Omega_{p}^{\prime*}}{A_{1}}\left[xy\Gamma_{43}+4y\left|\Omega_{p}^{\prime}\right|^{2}\right], \end{eqnarray}
\begin{eqnarray}
\rho^{ss}_{22} & = & \frac{4\left|\Omega_{p}^{\prime}\right|^{2}}{A_{1}}\left[\Gamma_{31}\Gamma_{42}+\Gamma_{32}\left(\Gamma_{42}+\Gamma_{43}\right)\right]\nonumber \\
 &  & \times\left[4\left|\Omega_{p}^{\prime}\right|^{2}+x\left(x+y\right)\right]\nonumber \\
 &  & +\frac{16\left|\Omega_{p}^{\prime}\right|^{2}}{A_{1}}\Gamma_{32}\left|\Omega_{s}^{\prime}\right|^{2}\left(\Gamma_{41}+\Gamma_{42}+x\right), \label{p1RES}\\
 %\rho^{ss}_{23} & = & 0, \\
%\rho^{ss}_{24} & = & 0,\\
 \rho^{ss}_{23} & = & \rho^{ss}_{24}=0, \\
%\end{eqnarray}
%\begin{eqnarray}
\rho^{ss}_{33} & = & \frac{4\Gamma_{21}\left|\Omega_{p}^{\prime}\right|^{2}}{A_{1}}\left[\begin{array}{c}
\Gamma_{43}\left(4\left|\Omega_{p}^{\prime}\right|^{2}+x\left(x+y\right)\right)\\
+4\left|\Omega_{s}^{\prime}\right|^{2}\left(\Gamma_{41}+\Gamma_{42}+x\right)
\end{array}\right], \label{p2RES}\\
%\end{eqnarray}
%\begin{eqnarray}
\rho^{ss}_{34} & = & -\frac{8ix\Gamma_{21}\left|\Omega_{p}^{\prime}\right|^{2}\Omega_{s}^{\prime*}}{A_{1}}\left(\Gamma_{41}+\Gamma_{42}+x\right),\\ \rho^{ss}_{44} & = & \frac{4x\Gamma_{21}\left|\Omega_{p}^{\prime}\right|^{2}}{A_{1}}\left(4\left|\Omega_{p}^{\prime}\right|^{2}+x\left(x+y\right)\right), %\end{eqnarray}
%\begin{eqnarray}
%\end{eqnarray}
%\begin{eqnarray}
%\rho^{ss}_{13} & = & -\frac{4\Omega_{s}^{\prime}\Gamma_{21}\Omega_{p}^{\prime*}}{A_{1}}\left[\begin{array}{c}
%\left(\Gamma_{41}+\Gamma_{42}+x\right)\left(4\left|\Omega_{s}^{\prime}\right|^{2}+xy\right)\\
%+4\left|\Omega_{p}^{\prime}\right|^{2}\left(\Gamma_{43}-x\right)+xy\Gamma_{43}
%\end{array}\right],\nonumber \\
%\end{eqnarray}
%\begin{eqnarray}
%\rho^{ss}_{12} & = &\rho^{ss}_{23}=\rho^{ss}_{24}= 0,
\end{eqnarray}
by solving the master equation~(\ref{eq:Mas_eq}). Other density matrix elements can be obtained by $\rho^{ss}_{mn}=\rho^{ss*}_{nm}$. Here
\begin{eqnarray}
\Gamma_{kj} & = & \sum_{\nu=1}^{3}\Gamma_{kj}^{\left(\nu\right)},\nonumber \\
x & = & \Gamma_{31}+\Gamma_{32},\nonumber \\
y & = & \Gamma_{41}+\Gamma_{42}+\Gamma_{43},
\end{eqnarray}
and the normalizing constant
\begin{widetext}
\begin{eqnarray}
A_{1} & = & \Gamma_{21}\left\{ 16\left|\Omega_{s}^{\prime}\right|^{4}\left(\Gamma_{41}+\Gamma_{42}+x\right)+xy\left[\left(x+y\right)\left(8\left|\Omega_{s}^{\prime}\right|^{2}+xy\right)-4\left|\Omega_{s}^{\prime}\right|^{2}\Gamma_{43}\right]\right\} \nonumber\\
 &  & +16\left|\Omega_{p}^{\prime}\right|^{4}\left[x\left(2\Gamma_{21}+\Gamma_{42}\right)+\left(\Gamma_{21}+\Gamma_{32}\right)\Gamma_{43}\right]\nonumber\\
 &  & +4\left|\Omega_{p}^{\prime}\right|^{2}\left\{ 4\left|\Omega_{s}^{\prime}\right|^{2}\left[\Gamma_{32}\left(\Gamma_{41}+\Gamma_{42}+x\right)+y\Gamma_{21}\right]+x\Gamma_{21}\left(2x^{2}+2xy+y^{2}\right)\right\} \nonumber\\
 &  & +4\left|\Omega_{p}^{\prime}\right|^{2}x\left(x+y\right)\left[x\Gamma_{42}+\left(\Gamma_{21}+\Gamma_{32}\right)\Gamma_{43}\right].
\end{eqnarray}
\end{widetext}

\subsection{Off resonance case}

In the large detuning regime, namely $E_{0}-\omega_{b}-\omega_{p}\gg\Omega_{p}c_{00}/2$,
and $E_{0}-\omega_{b}-2\omega_{c}-\omega_{s}\gg\Omega_{s}c_{02}/2$, and by requiring $|g_{2,4}|^{2}/\gamma_{3}^{2}\ll1$ and $|g_{0,2}|^{2}|g_{1,3}|^{2}/\gamma_{3}^{4}\ll1$, the system is truncated to a two-level system characterized by Hamiltonian~(\ref{eq:Heff}). According to this effective Hamiltonian~(\ref{eq:Heff}), by defining $\rho^{ss}_{n+1,m+1}\equiv\langle b,n|\rho^{ss}|b,m\rangle$ ($n,m=0,1,2$), we obtain the steady state solutions
\begin{eqnarray}
\rho^{ss}_{11} & = & 1-\frac{4\left|g_{0,2}\right|^{2}\left(\Gamma_{21}^{\left(3\right)}+\Gamma_{32}^{\left(3\right)}\right)}{A_{2}},\\ %\end{eqnarray}
%\begin{eqnarray}
\rho^{ss}_{22} & = & \frac{4\left|g_{0,2}\right|^{2}\Gamma_{32}^{\left(3\right)}}{A_{2}},\label{P1offres}\\
%\end{eqnarray}
%\begin{eqnarray}
\rho^{ss}_{33} & = & \frac{4\left|g_{0,2}\right|^{2}\Gamma_{21}^{\left(3\right)}}{A_{2}},\label{P2offres}
\end{eqnarray}
\begin{eqnarray}
\rho^{ss}_{12} & = & 0,\\
\rho^{ss}_{13} & = & \frac{2g_{0,2}\Gamma_{21}^{\left(3\right)}}{A_{2}}\left(i\Gamma_{31}^{\left(3\right)}+i\Gamma_{32}^{\left(3\right)}+2\Delta_{0}-2\Delta_{2}\right),\nonumber\\ \\
\rho^{ss}_{23} & = & 0,
%\rho^{ss}_{12} & = & \rho^{ss}_{23}=0,
\end{eqnarray}
by solving the master equation~(\ref{eq:Mas_eq}). Here the normalizing constant
is
\begin{eqnarray}
A_{2} & = & 4\left|g_{0,2}\right|^{2}\left(2\Gamma_{21}^{\left(3\right)}+\Gamma_{32}^{\left(3\right)}\right)\nonumber \\
 &  & +\Gamma_{21}^{\left(3\right)}\left[\left(\Gamma_{31}^{\left(3\right)}+\Gamma_{32}^{\left(3\right)}\right)^{2}+4\left(\Delta_{0}-\Delta_{2}\right)^{2}\right].\nonumber\\
\end{eqnarray}
\end{appendix}

\end{document}